\documentclass[12pt]{amsart}
\usepackage{amsmath,amscd,amssymb,latexsym}
\usepackage{graphicx}
\newtheorem{theorem}{Theorem}
\newtheorem{lemma}{Lemma}
\newtheorem{cor}{Corollary}

\def\a{{\alpha}}

\def\t{{\theta}}
\def\d{{\delta}}
\def\e{{\epsilon}}

\begin{document}
\title{A 2-chain can interlock with a $k$-chain}

\author{Julie Glass}
\address{Department of Mathematics \& Computer Science\\California State University Hayward\\Hayward, CA 94542}
\email{jglass@csuhayward.edu}

\author{Stefan Langerman}
\address{%
Universit\'e Libre de Bruxelles,
D\'epartement d'informatique,\\
ULB CP212, Bruxelles, Belgium}
\email{Stefan.Langerman@ulb.ac.be}

\author{Joseph O'Rourke}
\address{Department of Computer Science\\ Smith College\\ Northampton, MA 01063}
\email{orourke@cs.smith.edu}

\author{Jack Snoeyink}
\address{Dept. Comput. Sci., Univ. North Carolina,\\
Chapel Hill, NC 27599--3175, USA.}
\email{snoeyink@cs.unc.edu}

\author{Jianyuan K. Zhong}
\address{Department of Mathematics \& Statistics\\California State University Sacramento\\6000 J Street\\ Sacramento, CA 95819}
\email{kzhong@csus.edu}

\begin{abstract}
One of the open problems posed in~\cite{SoCG} is: what is the
minimal number $k$ such that an open, flexible $k$-chain can
interlock with a flexible $2$-chain? In this paper, we establish
the assumption behind this problem, that there is indeed some $k$
that achieves interlocking. We prove that a flexible 2-chain can
interlock with a flexible, open 16-chain. 
\end{abstract}
\maketitle

\section{Introduction}
A \emph{polygonal chain} (or just \emph{chain}) is a linkage of
rigid bars (line segments, edges) connected at their endpoints
(joints, vertices), which forms a simple path (an \emph{open
chain}) or a simple cycle (a \emph{closed chain}). A {\it folding}
of a chain is any  reconfiguration obtained by moving the vertices
so that the lengths of edges are preserved and the edges do not
intersect or pass through one another. The vertices act as
universal joints, so these are \emph{flexible chains}. If a
collection of chains cannot be separated by foldings, the chains
are said to be {\it interlocked}.

Interlocking of polygonal chains was studied in~\cite{DLOS,SoCG},
establishing a number of results regarding which collection of
chains can and cannot interlock.  One of the open problems posed
in~\cite{SoCG}
asked for the minimal $k$ such that a flexible open $k$-chain can
interlock with a flexible $2$-chain.  An unmentioned assumption
behind this open problem is that there is some $k$ that achieves
interlocking. It is this question we address here, showing that
$k=16$ suffices.

It was conjectured in~\cite{SoCG} that the minimal $k$ satisfies
$6 \le k \le 11$. This conjecture was based on a construction of
an 11-chain that likely does interlock with a 2-chain.
We employ some ideas from this construction in the example described
here, but for a 16-chain.
Our main contribution is a
proof that $k=16$
suffices.  It appears that using more bars makes it easier to
obtain a formal proof of interlockedness.

Results from~\cite{SoCG} include:
\begin{enumerate}
\item Two open 3-chains cannot interlock. 
\item No collection of 2-chains can interlock. 
\item A flexible open 3-chain can interlock
with a flexible open 4-chain.
\end{enumerate}

This third result is crucial to the construction we present, which
establishes our main theorem, that
a 2-chain can interlock a 16-chain (Theorem~\ref{main.theorem} below.)

\section{Idea of Proof}
\begin{figure}[htbp]
\centering
\includegraphics[width=0.8\linewidth]{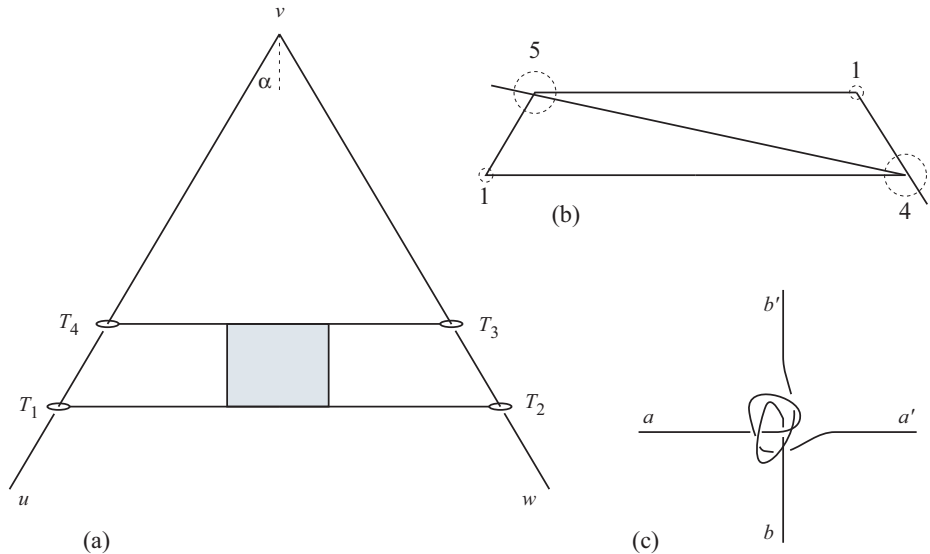}
\caption{(a) A rigid trapezoid with rings would interlock with a
2-chain; (b)~An open chain that simulates a rigid trapezoid;
(b)~Fixing a crossing of $aa'$ with $bb'$.} 
\label{Trapezoid.idea}
\end{figure}

We first sketch the main idea of the proof. If we could build a
rigid trapezoid with small rings at its four vertices
$(T_1, T_2, T_3, T_4)$, this could
interlock with a 2-chain, as illustrated in
Figure~\ref{Trapezoid.idea}(a). For then pulling vertex $v$ of the
2-chain away from the trapezoid would necessarily diminish the
half apex angle $\a$, and pushing $v$ down toward the trapezoid
would increase $\a$. But the only slack provided for $\a$ is that
determined by the diameter of the rings. We make as our subgoal,
then, building such a trapezoid.

We can construct a trapezoid with four links, and rigidify it with
two crossing diagonal links.  In fact, only one diagonal is
necessary to rigidify a trapezoid in the plane, but clearly a
single diagonal leaves the freedom to fold along that diagonal in
3D.  This freedom will be removed by the interlocked 2-chain,
however, so a single diagonal suffices. To create this rigidified
trapezoid with a single open chain, we need to employ $5$ links,
as shown in Figure~\ref{Trapezoid.idea}(b). But this will only be
rigid if the links that meet at the two vertices incident to the
diagonal are truly ``pinned'' there.  In general we want to take
one subchain $aa'$ and pin its crossing with another subchain
$bb'$ to some small region of space. See
Figure~\ref{Trapezoid.idea}(c) for the idea.

This pinning can be achieved by the ``$3/4$-tangle'' interlocking from~
\cite{SoCG}, result~(3) above. So the idea is replace the two critical crossings
with a small copy of this configuration. This can be
accomplished with $7$ links per $3/4$-tangle, but
sharing with the incident incoming and outgoing trapezoid links potentially reduces
the number of links needed per tangle.  We have achieved
$5$ links at one tangle and $4$ at the other.
The other two
vertices of the trapezoid need to simulate the rings in
Figure~\ref{Trapezoid.idea}(a), and this can be accomplished with
one extra link per vertex. Together with the $5$ links for the
main trapezoid skeleton, we employ a total of $5+ (5+4+1+1) = 16$
links.

\section{A 2-chain can interlock an open 16-chain}
\subsection{Open flexible 3- and 4-chains can interlock}

\begin{figure}[htbp]
\centering
\includegraphics[width=0.95\linewidth]{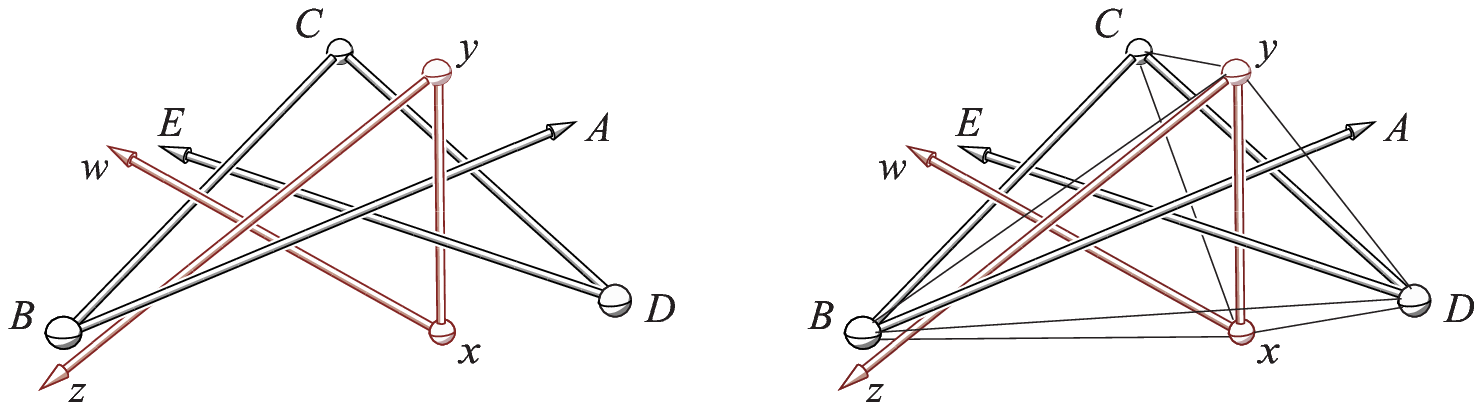}
\caption{Fig.~6 from \protect\cite{SoCG}.}
\label{link34}
\end{figure}

It was proved that open flexible 3- and 4-chains can interlock in
\cite{SoCG}. 
The construction, which we call a \emph{3/4-tangle},
is repeated in Figure~\ref{link34}.

It was proved in Theorem~11 of~\cite{SoCG} that the convex hull
$CH(B,C,D,x,y)$ of the 
joints $B$, $C$, $D$, $x$, and $y$
does not change.

We first establish bounds on how far the vertices of the construction
can move.
Let $BC=CD=xy=1$ unit, and end bars $AB=DE=xw=yz=3$ units.

\begin{lemma}
Let $P$ be the midpoint of $xy$.  Then in any folding of the
interlocked 3- and 4- chain: (1) The distance between $P$ and the
endpoints $w$, $z$ of the 3-chain can be no more than $3.5$ units,
(2) The distance between $P$ and joints $B$, $C$, $D$, $x$, and
$y$ can be no more than $2.5$ units, and (3) The distance between
$P$ and the endpoints $A$, $E$ of the 4-chain can be no more than
$5.5$ units.
\end{lemma}
\begin{proof}
(1) Since $P$ is the midpoint of bar $xy$, $x$ and $y$ are exactly
$0.5$ units away from $P$. The joints $w$, $x$ and $P$ form a
triangle, by the triangle inequality $Pw<Px+xw=0.5+3=3.5$ units;
similarly, $Pz<3.5$.

(2) We now prove that the distance between $P$ and the joints $B$,
$C$, $D$, $x$, and $y$ can be no more than $2.5$ units. In the
convex hull $CH(B,C,D,x,y)$, bar $xy$ pierces $\triangle BCD$,
where $B$ and $D$ can be imagined to be connected by a rubber
band, then $BD<BC+CD=2$. We observe that: (i) any two points
inside $\triangle BCD$ or on the boundary $BC, CD, BD$ are less
than $2$ units apart, and (ii) the distance between the midpoint
$P$ and the plane determined by $B, C, D$ must be less than $0.5$
units. From the fact that bar $xy$ pierces $\triangle BCD$, the
distance between $P$ and any point on or inside $\triangle BCD$ is
less than $2+0.5=2.5$ units. Since $P$ is the midpoint of bar
$xy$, $x$ and $y$ are exactly $0.5$ units away from $P$.
Therefore, $P$ and the joints $B$, $C$, $D$, $x$, and $y$ can be
no more than $2.5$ units as claimed.

(3) Finally, by the triangle inequality $PA<PB+AB<2.5+3=5.5$
units; similarly, $PE<5.5$.
\end{proof}

For $\e>0$, choosing $BC=CD=xy=\frac{1}{6}\e$, and end bars
$Ab=DE=xw=yz=\frac{1}{2}\e$ yields the following:
\begin{cor}
In the above interlocked 3- and 4-chains, let $P$ be the midpoint
of $xy$, then all joints $B,C,D,x,y$ and endpoints $A, E, w, z$
stay inside the $\e$-ball centered at $P$.
\end{cor}

\subsection{A 2-chain can interlock an open 16-chain}
\begin{figure}[htbp]
\centering
\includegraphics[width=0.8\linewidth]{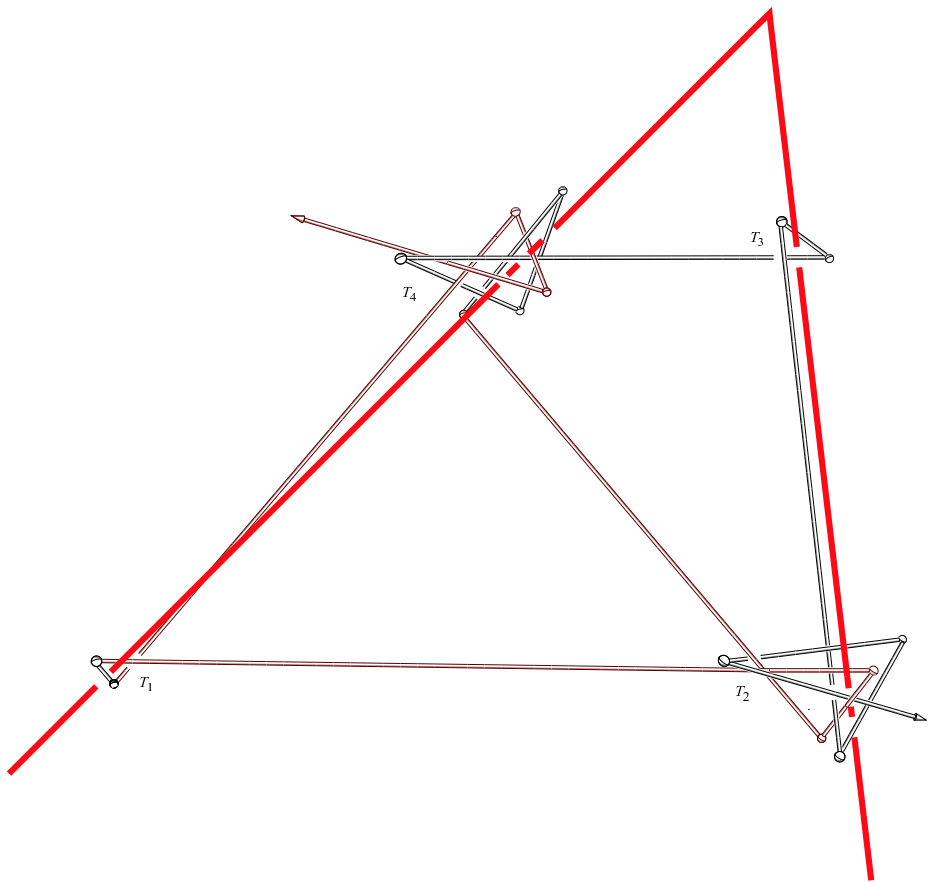}
\caption{An open 16-chain forming a nearly rigid trapezoid.}
\label{16chain}
\end{figure}

Take two 3/4-tangles, where all joints
and end points of the pair stay within an $\e$-ball centered at
the midpoint of the middle link of the 3-chain. Position the tangles
as two of the ``vertices'' of a trapezoid with the links arranged
as shown in Figure~\ref{16chain}.
This design follows 
Figure~\ref{Trapezoid.idea}(b) in spirit, but varies the connections at
the diagonal endpoints to
increase link sharing.  
The lower right vertex achieves maximum sharing,
in that all three incident trapezoid edges are shared with links
of the $3/4$-tangle.  The upper left vertex shares two incident links.
We extend the
first and last links of the trapezoid chain to be very long so that the end vertices
of the chain are well exterior to any of the $\e$-balls.

\subsubsection{2-chain Through Trapezoid Jag Corners}
Call the simple structure at the other two corners
\emph{jag loops}.
These corners also can be assured to remain 
in an $\e$-ball simply by making the extra link
length $\e$.  Thus we have that all corners of the trapezoid stay
within $\e$-balls. 

\begin{figure}[htbp]
\centering
\includegraphics[width=0.5\linewidth]{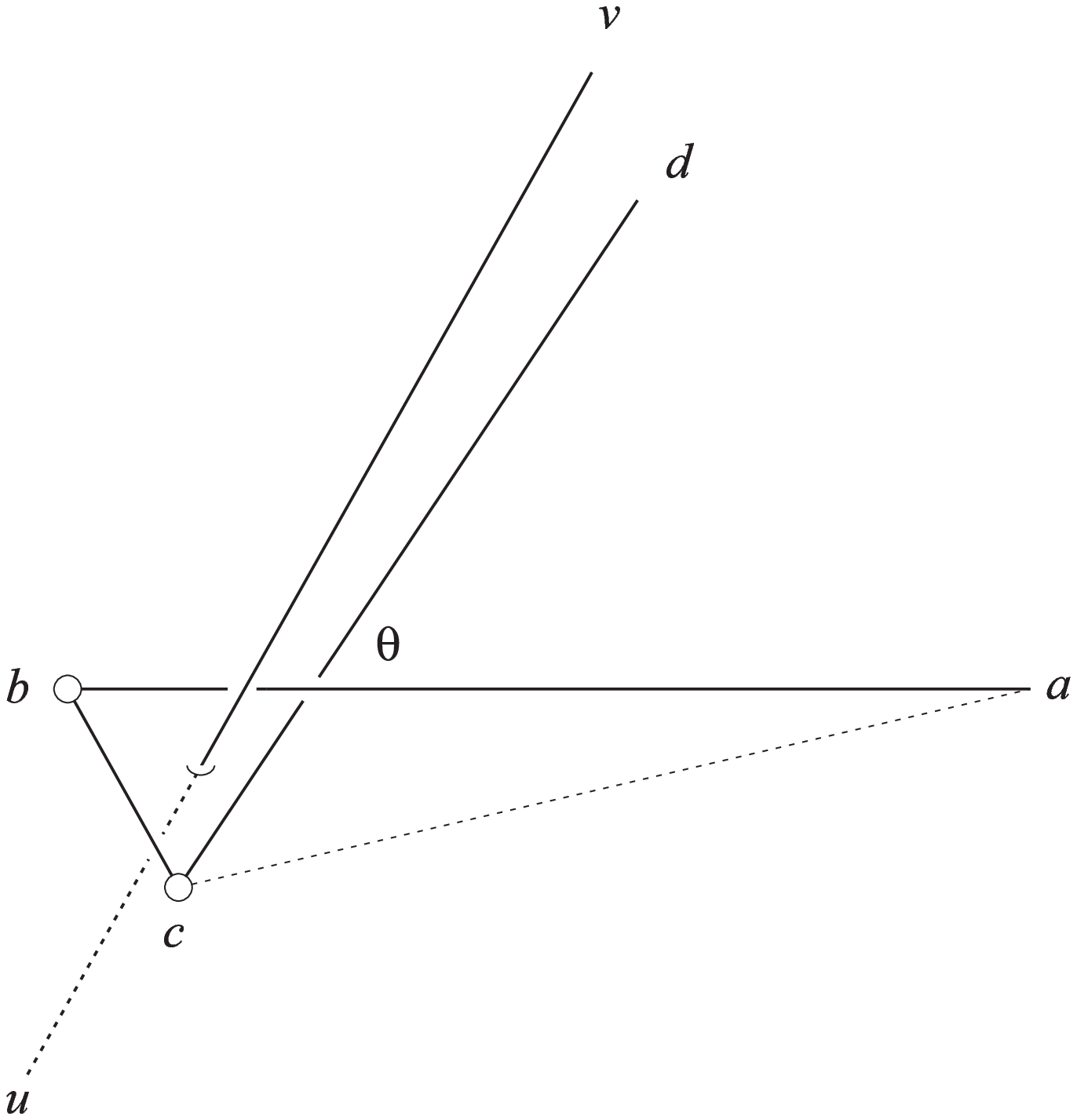}
\caption{A 1-link ``jag.''}
\label{1-link.jag}
\end{figure}

We first argue that the jag loop ``grips'' the 2-chain link through
it, under the assumption of near rigidity of the trapezoid.
Let $(u,v,w)$ be the 2-link chain,
and let $(a,b,c,d)$ be the vertices constituting a 1-link jag
at a corner of the trapezoid.
The short link of the jag is $bc$.
The near-rigidity of the trapezoid permits us to
take $ab$ to be roughly horizontal (the base of the trapezoid)
and $cd$ to be roughly at angle $\t$ with respect to the base 
(the angle at a base corner of the trapezoid).
The link $uv$ is nearly parallel to $de$, and is woven through
the jag as illustrated 
in Figure~\ref{1-link.jag}.
The words ``roughly'' and ``nearly'' here are intended
as shorthand for ``approaches, as $\e \rightarrow 0$.''
\begin{lemma}
The plane containing $\triangle abc$ continues to separate $v$ above from
$u$ below (where ``above'' is determined by the counterclockwise
ordering of $a,b,c$)
under all nonintersecting foldings of the chains.
\end{lemma}
\begin{proof}
We argue that $uv$ continues to properly pierce $\triangle abc$
under all foldings, from which it follows that the initial separating
property is maintained.
The overall structure of the trapezoid prevents $uv$ from moving directly through
$\triangle abc$: neither $v$ nor $u$ can get close to the triangle.
So the only way the piercing could end is if $uv$ passes through
a side of $\triangle abc$.
Two of these sides---$ab$ and $bc$---are links, and avoiding intersection
prevents passage through those.
Thus $uv$ would have to pass through $ac$, which is not a link.
However, to do this, we now argue it would have to pass through the 
link $cd$.

The gap between $ab$ and $cd$ is at most $|bc|=\e$.
$uv$ must pass through this gap to ``escape'' and pass through the segment $ac$.
Because $|uv| \gg \e$, $uv$ must turn ``sideways'' to pass through it.
More precisely, let $Q$ be a plane parallel to $ab$ and $cd$ and midway between
them, i.e., $Q$ passes through the midpoint of the gap.
$uv$ must align to lie nearly in $Q$ to pass through the gap.
Because $uv$ is on the ``wrong side'' of $ab$,
there are only two ways $uv$ can reach $Q$:
either to align roughly parallel to $ab$, or to
align roughly parallel to $cd$.
In either case, it would then be possible to
pass $uv$ through the gap, by keeping it close to the long link to which
it is nearly parallel.
However, the first alignment places $uv$ at an angle near $\t$ with respect
to $cd$; but it must be nearly parallel to $cd$.
The second alignment requires flipping $uv$ around so that $u$ is above $v$
in the view shown in the figure, in order to get on the other side of $ab$.
But this then makes $uv$ approximately antiparallel to $cd$,
rather than nearly parallel as it must be.
Thus the only escape route is impossible, and
$uv$ maintains its piercing of $\triangle abc$.
\end{proof}

\begin{cor}
The 1-link jag interlocks with $uv$, under the constraints imposed by the nearly rigid trapezoid.
\label{cor.jag}
\end{cor}

\subsubsection{2-chain Through Trapezoid Tangle Corners}
Next we argue that the link $uv$ can thread through the corner $T_4$
of the trapezoid so that it is ``gripped'' by the $3/4$-tangle there.
Note that the $(T_1,T_4)$ trapezoid link connects to the 3-chain
at $T_4$, which is itself just a jag loop.
But $uv$ cannot thread properly through both jag loops on either
end of the $(T_1,T_4)$ link.  So instead we thread $uv$ through the 4-chain at $T_4$.

\begin{figure}[htbp]
\centering
\includegraphics[width=0.3\linewidth]{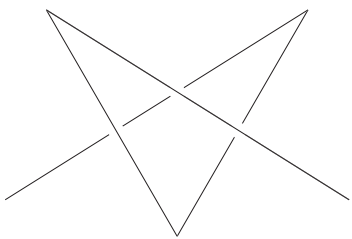}
\caption{A 4-chain, part of a 3/4-tangle, can be viewed as two jag loops.}
\label{4-chain}
\end{figure}

Now, a 4-chain can be viewed as two jag loops; see Figure~\ref{4-chain}.
Moreover, the 4-chain and 3-chain participating in a $3/4$-tangle
can be viewed as each lying nearly in planes that are twisted with
respect to one another.  So we chooose to twist the 4-chain at $T_4$
so that $uv$ threads properly through one of its two jag loops.
Similarly, the link $vw$ threads from the jag at $T_3$ through the
3-chain at $T_2$ 
(and not through the 4-chain to which $(T_4,T_2)$ is connected).

Applying Corollary~\ref{cor.jag} to guarantee interlocking yields:
\begin{lemma}
The 2-chain links, when threaded as just described,
are interlocked with the $3/4$-tangles, 
under the constraints of a nearly rigid trapezoid.
\label{tangle.thread}
\end{lemma}

We should mention that the foregoing argument would be unnecessary if
we had instead used a 2-link jags at $T_1$ and $T_3$, which would
give freedom to position the jag to permit piercing the tangles
however desired (and which would lead to an 18-link interlocking chain).

Finally, there is more than enough flexibility in the design to 
ensure that $uv$ and $vw$ can indeed share the same 2-chain apex $v$.

\subsubsection{Apex $v$ Cannot Move Far}
Thus
the 2-chain $(u,v,w)$ cannot slide free of any of the trapezoid corners unless
one of its vertices enters the $\e$-ball containing the corner.
We argue below that this cannot occur.  We start with a simple preliminary
lemma.

\begin{figure}[htbp]
\centering
\includegraphics[width=0.6\linewidth]{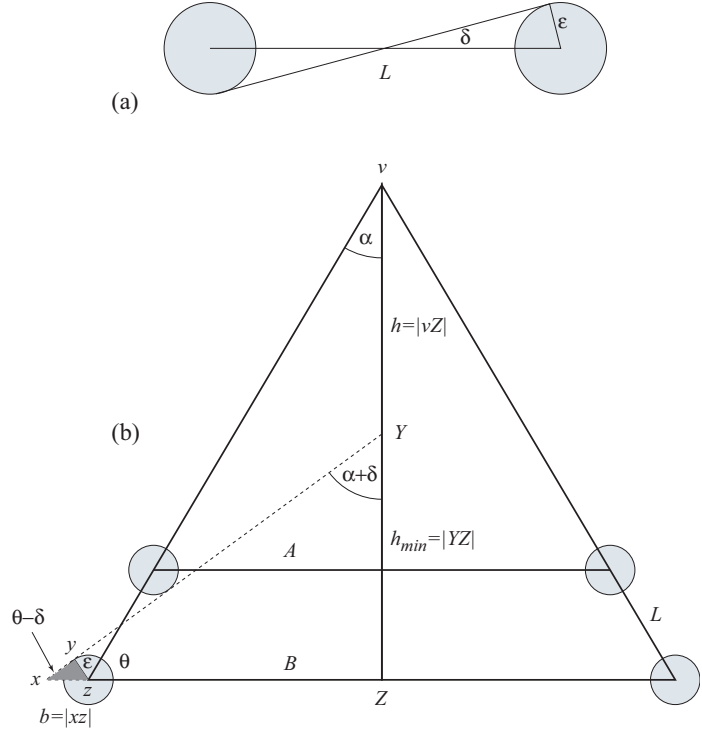}
\caption{Trapezoid Lemma: (a) Line through two disks deviates at
most $\d$; (b) Trapezoid structure, with $h_{min}$ computation
illustrated.}
\label{Trapezoid.Lemma}
\end{figure}

\begin{lemma}
When $\e$ is sufficiently small, a line piercing two disks of
radius $\e$ can angularly deviate from the line connecting the
disk centers at most $\d \le  2 \e / L$, where $L$ is the distance
between the disk centers.
\end{lemma}
\begin{proof}
Figure~\ref{Trapezoid.Lemma} illustrates the largest angle $\d$,
$(\frac{1}{2}) L \sin \d = \e$, so $\sin \d = 2 \e /L$, and the
claim follows from the fact that 
$\displaystyle{\lim_{x\to0}\frac{\sin x}{x}=1}$.
\end{proof}

Let the trapezoid have base of length $2B$, side length $L$, and
base angle $\t$. Let the triangle determined by the trapezoid have
height $h$ and half-angle $\a$ at the apex, so $\tan \t = h/B$, or
$h = B \tan \t$. See Figure~\ref{Trapezoid.Lemma}(b).
The following lemma captures the key constraint on motion of the 2-link.

\begin{lemma}
If the sides of the trapezoid pass through the $\e$-disks
illustrated, then the height of the triangle approaches $h$ as
$\e \rightarrow 0$.
\end{lemma}
\begin{proof}
$h_{min}$ occurs with a triangle apex angle of $\a + \d$ and a
base angle of $\t -\d$. Let $b$ be the amount by which $B$ is
lengthened.  In $\triangle xyz$, $b =
\frac{\epsilon}{\sin(\theta - \delta)}$, and in $\triangle xYZ$,
$\tan(\theta - \delta) = \frac{h_{min}}{B + b}$.  Thus
we have that 
$$h_{min} = \left(B+ \frac{\epsilon}{\sin(\theta -
\delta)}\right) \tan(\theta - \delta) = B\tan(\theta - \delta) +
\frac{\epsilon}{\cos(\theta - \delta)}.$$  
Thus $h_{min}$ is
continuous near $\epsilon = 0$.  Also, if $\epsilon \rightarrow 0$
then $\delta \rightarrow 0$ since $\delta \leq
\frac{2\epsilon}{L}$. Therefore  $\lim_{\epsilon \rightarrow
0}h_{min} = B\tan{\theta} = h$ since $\tan(\theta) = h/B$.

For $h_{max}$ all the signs reverse to yield that
$\lim_{\epsilon \rightarrow 0}h_{max} = B\tan{\theta} = h.$

We conclude that the height of the triangle
approaches $h$ as $\epsilon$ approaches 0 as desired.
\end{proof}


\subsubsection{Main Theorem}
We connect 3D to 2D via the plane determined by the 2-link in the proof
of the main theorem below.

\begin{theorem}
The 2-link chain is interlocked with the 16-link trapezoid chain.
\label{main.theorem}
\end{theorem}
\begin{proof}
Let $H$ be the plane containing the 2-link chain. We know that the
links of the 2-chain must pass through $\e$-balls around the four
vertices of the trapezoid. $H$ meets these balls in disks each of
radius $\le \e$. The Trapezoid Lemma shows that the height of the
triangle approaches $h$ as $\e$ approaches 0. Thus, by choosing
$\e$ small enough, we limit the amount that the apex $v$ of the
2-link chain can be separated from or pushed toward the trapezoid
to any desired amount. 

We previously established 
(in Corollary~\ref{cor.jag} and Lemma~\ref{tangle.thread})
that the 2-chain links are interlocked
with the $3/4$-tangles and jag loops through which they pass,
under the assumption that the trapezoid is nearly rigid.
The near-rigidity of the trapezoid could only be destroyed by
a 2-chain link escaping from one of the jag loops through which it
is threaded.  But up until the time of this first escape, the
trapezoid is nearly rigid; and so there can be no first escape.

Thus, choosing $\e$ small enough to prevent any of the
vertices of the 2-link chain from entering the $\e$-balls ensures
that the 2-link chain is interlocked with the trapezoid chain.
\end{proof}

\section{Discussion}

We do not believe that $k=16$ is minimal. 
We have designed two different 11-chains both of which appear
to interlock with a 2-chain.
However, both are
based on a triangular skeleton rather than on a
trapezoidal skeleton, and place the apex $v$ of the 2-chain
close to the 11-chain. 
It seems it will require a different proof
technique to establish interlocking, for the simplicity of the
proof presented here relies on the vertices of the 2-chain remaining far from
the entangling chain.

Another direction to explore is closed chains, for which it is
reasonable to expect fewer links.
Replacing the $3/4$-tangles with 
``knitting needles'' configurations~\cite{CJ}\cite{knitting}
produces a closed chain that appears interlocked,
but we have not determined the minimum number of links that can achieve this.

\subsection*{Acknowledgement}
We thank Erik Demaine for discussions throughout this work.
We thank the participants of the DIMACS Reconnect Workshop held
at St. Mary's College in July 2004 for helpful discussions.
JOR acknowledges support from NSF DTS award
DUE-0123154.

\end{document}